\documentclass[12pt]{iopart}
\usepackage[german,american,english]{babel}
\usepackage{graphicx}% Include figure files
\usepackage{graphics}% Include figure files
\usepackage{dcolumn}% Align table columns on decimal point
\usepackage{multirow}
\usepackage{bm}% bold math
\usepackage{latexsym}
\usepackage{iopams}
\usepackage{layout}
\usepackage{verbatim}
\usepackage{epsfig}
\usepackage{graphicx}
\usepackage{lipsum}
\usepackage{color}
\usepackage{cite}
\newcommand{\bea}{\begin{eqnarray*}}
	\newcommand{\eea}{\end{eqnarray*}}
\newcommand{\bne}{\begin{equation*}}
\newcommand{\ede}{\end{equation*}}

\newcommand{\bnen}{\begin{eqnarray}}
\newcommand{\eden}{\end{eqnarray}}
\newcommand{\bean}{\begin{eqnarray}}
\newcommand{\eean}{\end{eqnarray}}
\newcommand{\bsen}{\begin{subequations}}
	\newcommand{\esen}{\end{subequations}}

\newcommand{\dps}{\displaystyle}

\newcommand{\om}{\iffalse}
\newcommand{\pd}[2]{\frac{\partial #1}{\partial #2}}

\newcommand{\ba}{\arraycolsep 0.3ex \begin{array}{rl}}
\newcommand{\ea}{\end{array}}
\newcommand{\bc}{\begin{cases}}
\newcommand{\ec}{\end{cases}}

\newcommand{\bna}{\begin{array}}
	\newcommand{\eda}{\end{array}}
\newcommand{\bnm}{\begin{enumerate}}
	\newcommand{\edm}{\end{enumerate}}

\usepackage[utf8]{inputenc}

\begin{document}

\title[Coherent backscattering in the topological Hall effect]{Coherent backscattering in the topological Hall effect}
\author{Hong Liu$^{1,2,3}$, Rhonald Burgos Atencia$^{3,4, 5}$, Nikhil Medhekar$^{1,2}$ and  Dimitrie Culcer$^{3,4}$ }
\address{$^1$ School of Physics and Astronomy, Monash University, Victoria 3800, Australia}
\address{$^2$ ARC Centre of Excellence in Future Low-Energy Electronics Technologies, Monash University, Victoria 3800, Australia}
\address{$^3$School of Physics, The University of New South Wales, Sydney 2052, Australia}
\address{$^4$ARC Centre of Excellence in Low-Energy Electronics Technologies, UNSW Node, The University of New South Wales, Sydney 2052, Australia}
\address{$^5$ Facultad de Ingenier\'ias, Departamento de Ciencias B\'asicas, Universidad del Sin\'u, Cra.1{\rm w} No.38-153, 4536534,
Monter\'ia, C\'ordoba 230002, Colombia}
\ead{hong.liu1@monash.edu {\rm and} d.culcer@unsw.edu.au}
\begin{indented}
\item[]November 2022
\end{indented}
\begin{abstract}
The mutual interplay between electron transport and magnetism has attracted considerable attention in recent years, primarily motivated by strategies to manipulate magnetic degrees of freedom electrically, such as spin-orbit torques and domain wall motion. Within this field the topological Hall effect, which originates from scalar spin chirality, is an example of inter-band quantum coherence induced by real-space inhomogeneous magnetic textures, and its magnitude depends on the winding number and chiral spin features that establish the total topological charge of the system. Remarkably, in the two decades since its discovery, there has been no research on the quantum correction to the topological Hall effect. Here we will show that, unlike the ordinary Hall effect, the inhomogeneous magnetization arising from the spin texture will give additional scattering terms in the kinetic equation, which result in a quantum correction to the topological Hall resistivity. We focus on 2D systems, where weak localization is strongest, and determine the complicated gradient corrections to the Cooperon and kinetic equation. Whereas the weak localization correction to the topological Hall effect is not large in currently known materials, we show that it is experimentally observable in dilute magnetic semiconductors. Our theoretical results will stimulate experiments on the topological Hall effect and fill the theoretical knowledge gap on weak localization corrections to transverse transport.
\end{abstract}

\submitto{Materials for Quantum Technology}
\maketitle

% We need at least 10-20 references in the first paragraph, and they should cover everything that is mentioned - SOT, hard disks, logic and memory devices, vortices, merons, etc. (done)

\section{Introduction}\label{Intro} 

Classical information technology relies heavily on electron transport and magnetism, with semiconductors and ferromagnets playing complementary roles in information processing and storage \cite{Spin-torque-Rev}. Magnetism has been most successful in high-density memory applications \cite{High-density-MR-RAM}, such as hard disks \cite{Hard-disk-APL,Hard-disk}, while considerable research is being devoted to switching magnetic memory components by electrical currents via spin-orbit torques\cite{Roadmap-SOT,SOT-Materials,SOT-Topological-Materials,SOT-RMP-2019,AHE-CULCER2022,Dimi-SOT-TI,DFT-SOT-book,SOT-APR}. % References here from Jimmy's papers, my review on topological materials SOT, my review on AHE arXiv:2204.02434. (done)
The latter effort has been galvanised by the rise of topological materials, where spin-orbit interactions are often strong, and room-temperature magnetization switching has been accomplished by means of an electrical current \cite{Room-T-M-Switch}. % Hyunsoo (done)
At the same time the discovery of topological magnetic materials has generated considerable interest in topologically protected spin textures, including vortices, merons and skyrmions with potential application in next-generation logic and memory devices \cite{Merion-SK-Vor-PRB,Merons-Nat-Comm}.

Electrical transport in media hosting skyrmions and chiral spin textures is influenced by the exchange interaction between itinerant carriers and localized spins forming such textures. A famous example of this interplay is the topological Hall effect \cite{Bruno-THE-PRL-2004, Skymions-Nagaosa,AHE-Texture-TI,Giant-THE-NdMn2Ge2,THE-Small,AHE-kagome-PRR,Monopole-PRL,PhysRevLett.125.247201,THE-Hopfions}. Unlike the anomalous Hall effect, which relies on the spin-orbit interaction \cite{AHE-CULCER2022}, the topological Hall effect stems from the inhomogeneous spin texture that yields a fictitious magnetic field, not requiring any spin–orbit coupling. Stabilization and dynamics of different chiral spin textures depend strongly on their topological properties \cite{High-density-Neel-npj,PhysRevResearch.3.023109,Magnetic-skymions-nat}. The effect was first discussed in two-dimensional diluted magnetic semiconductors \cite{Bruno-THE-PRL-2004} but has since grown into a distinct research area in its own right. Then topological Hall effect has been studied theoretically in references \cite{WeakCoupling-THE-RPB-2019,ZHANG2021167700,THE-simple-model-PRB-2006,PhysRevB.98.195439,PhysRevB.97.134401,THE-W-S-theory,PhysRevB.98.214407,Denisov-Nonadiabati-PRL-2016,Skyrmion-soccer}, where spin-flip scattering \cite{PhysRevB.97.134401,Denisov-Nonadiabati-PRL-2016} and the evolution from strong exchange coupling to weak exchange coupling \cite{THE-weak-coupling-PRB} are explicitly considered. To date the topological Hall effect has been experimentally observed in a vast range of materials
\cite{Di-Wu-THE-NL-2019,Lan-Nat-comm,THE-exp-PRB-2021,PhysRevB.103.L241112,PhysRevMaterials.5.034405,THE-signatures-ACSNano,NL-THE-TI}, summarized in a recent review \cite{WANG-THE-review}. They include ferromagnetic semiconductors, magnetic Kagome lattices, transition-metal oxides, SrRuO$_3$-based materials, transition metal compounds, Heusler compounds, and magnetic topological insulators, among others \cite{Skyrmion-bubble-PRR, PhysRevMaterials.4.054414,Intrinsic-magnetic-TI,Topological-Phase-transiton-PRB}. Topological spin textures that might produce topological Hall effect include Bloch-type skyrmions induced by bulk Dzyaloshinskii–Moriya interaction (DMI), N\'eel-type skyrmions mainly associated with interfacial DMI, scalar spin chirality generally originated from geometrical frustration interaction, and noncollinear magnetism and bubble caused by the competition of various magnetic interactions including magnetocrystalline anisotropy, demagnetization and exchange interaction \cite{WANG-THE-review}.

Transport studies have overwhelmingly focused on the classical regime. Theoretical studies to date have primarily relied on classical transport theory ($\hbar/\epsilon_{\rm F}\tau\ll1$ with $\epsilon_{\rm F}$ Fermi energy, $\tau$ momentum relaxation time and $\hbar$ Planck constant). Nevertheless, besides classical transport, quantum coherence modifies electronic transport properties significantly. The weak localization correction to the conductivity arises as a result of the quantum interference between closed, time-reversed loops that circle regions in which one or more impurities are present, and is noticeable when the electron mean free path is much shorter than the phase coherence length. Weak localization is strongly affected both by spin-orbit interactions in the band structure and by the form of the impurity potential, as well as by external inhomogeneous magnetic fields and by domain walls \cite{Nat-WL-GaMnAs, PhysRevLett.78.3773, Daniel-WL-PRB-1993, Yuli-Domain-Wall-WL-PRL}. In this context, the effect on weak localization of inhomogeneous magnetic structures in real space has received almost no attention, despite the inexorable rise of magnetic systems with topological spin textures such as skyrmions. One notable exception is a recent study of weak localization in longitudinal transport due to scattering off chiral spin textures  \cite{Golub-WL-SpinTexture-PRB-2019}. To date no studies have considered weak localization in the Hall conductivity of topological spin textures, presumably because the effect is expected to vanish in ordinary Hall transport \cite{Altshuler-Hall-2DEG-PRB}, though it is believed to play a part in the anomalous Hall effect~\cite{WL-AHE-Bruno}. The fundamental scientific questions in this context are: \textit{What is the effect of inhomogeneous topological magnetic textures on coherent backscattering? Can there be a weak localization correction to Hall transport?} The resolution of these questions requires a complex kinetic equation, in which the transport theory needs to be augmented by a series of non-trivial gradient corrections. 

We seek to answer these fundamental questions in the present work by investigating theoretically the weak localization correction to the topological Hall effect in 2D systems with chiral spin textures. We derive a generalized quantum kinetic equation that captures the effect of inhomogeneous magnetic textures and provide a complete quantum kinetic framework for describing the topological Hall effect including corrections up to the second order in the spatial gradient. We show that the topological Hall effect is a manifestation of inter-band quantum coherence induced by a real-space magnetic texture \cite{Sekine-PRB}, and accounting for the interplay between diagonal and off-diagonal density matrix response is vital in order to capture the underlying physics and determine the correct result both in the Born approximation and for coherent backscattering. We restrict ourselves to scalar impurity scattering, noting that spin-flip scattering can straightforwardly be included along the lines of reference \cite{Golub-WL-SpinTexture-PRB-2019}. This however will only increase the complexity of the theory by modifying the Cooperon quantitatively without bringing about any qualitative changes in our results. 

We concentrate on a simple model that captures inter-band coherence between spin up and spin down bands, namely that of a two-dimensional electron gas with an inhomogenous magnetization. Due to the interplay of magnetic inhomogeneity and inter-band coherence we find a series of additional scattering terms in the kinetic equation stemming from the gradient of the magnetization, which give rise to a weak localization correction in the transverse transport direction. Since the interference effects leading to weak localization disappear in weak external magnetic fields, these corrections can typically be identified straightforwardly in experiment, and are frequently used to characterize samples, in particular transport in novel materials. They provide valuable information about the system, such as symmetries of the system, the phase coherence length, and the quasiparticle mass \cite{Weizhe-WL-Material}. The weak localization correction is somewhat smaller than the topological Hall effect itself, but it is observable in systems with sizable topological Hall resistivities, hence we expect our result to stimulate experiments.

Our paper is organized as follows. Section~\ref{Method} presents a comprehensive discussion of our method. We derive a general quantum kinetic equation that accounts both homogeneous and inhomogenous magnetization in section~\ref{Kin-Equation}. Then the gradient correction to the scattering term arising from the inhomogenous magnetization is shown in section~\ref{Gradient-J}. 
In sections~\ref{THE} and~\ref{WLxx}, we solve the quantum kinetic equation without gradient correction in scattering term, which will reproduce the topological Hall effect in transverse direction and the weak localization in the longitudinal direction. 
In section~\ref{WL-THE}, we calculated the weak localization correction to the topological Hall effect by taking the additional gradient scattering term in the quantum kinetic equation. In section~\ref{Result}, we provide a brief overview and physical explanation of our main results, then apply our model to various skyrmion systems and discuss experimental observation. In section~\ref{Conc} we summarize our findings and outlook.

\section{Theoretical Formalism}\label{Method}

\subsection{Quantum kinetic equation}\label{Kin-Equation}
The total Hamiltonian of the system is
\begin{eqnarray}
    H=H_0+U({\bm r})+H_{\bm E}.
\end{eqnarray}
Here $H_0$ is the band Hamiltonian describing a two dimensional electron system with an inhomogeneous magnetization. The band Hamiltonian can be further decomposed into $H_0=h_{\bm k}+H_{\rm M}({\bm r})$, where the kinetic energy $h_{\bm k}=\frac{\hbar^2k^2}{2m}$, with $m$ effective mass and $\hbar$ the Planck constant. The spatially varying magnetization gives rise to a Zeeman-like splitting $H_{\rm M}({\bm r})={\bm \sigma}\cdot {\bm M}({\bm r})$, where $M$ has units of energy. For simplicity, we assume the local magnetization ${\bm M}({\bm r})=M{\bm n}({\bm r})$, and that the 3D unit vector ${\bm n}({\bm r})=(\sin\theta\cos\phi_r,\sin\theta\sin\phi_r,\cos\theta)$ is a slowly varying function of the spatial coordinates, where $\cos\theta=\epsilon_k/\sqrt{\epsilon^2_k+M^2}$ and $\sin\theta=M/\sqrt{\epsilon^2_k+M^2}$. The phase $\phi_r$ is dependent on location.
The conduction electrons are considered to be non-interacting, but are scattered by spin-independent impurities. The disorder potential consists of uncorrelated short-range scalar impurities. The total impurity potential is  $U({\bm r})=\sum_{I}U({\bm r}-{\bm R}_{I})$, where $\sum_{I}$ is a summation over all impurities and ${\bm R}_I$ is the coordinate of impurity $I$. The reciprocal-space matrix element of a short-range impurity potential matrix element is $U_{{\bm k}{\bm k}'}\equiv u$. The customary solution to dealing with disorder is performing an average over all the possible impurity configurations, whereupon the total impurity potential in reciprocal space $U_{{\bm k}{\bm k}'}$ becomes $\langle U_{{\bm k}{\bm k}'}\rangle=0$ and $\langle U_{{\bm k}{\bm k}'}U_{{\bm k}_1{\bm k}'_1}\rangle=n_{\rm i} u^2$. $n_{\rm i}$ represents the impurity concentration. $H_{\bm E}$ is the external electric driving field. 

Our discussion is based on the density matrix transport theory, where central quantity is density matrix $\hat{\rho}$ in the quantum Liouville equation
\begin{eqnarray}
\frac{\partial \hat{\rho}}{{\rm d}t}+\frac{\rm i}{\hbar}[H,\hat{\rho}]=0.
\end{eqnarray}
We decompose the density matrix $\hat{\rho}=f +g$, where $f$ is disorder-averaged density matrix and $g$ is the fluctuating part, which is integrated over within a chosen approximation scheme that determines the effective scattering term, discussed in detail below. We obtain the kinetic equation for the disorder-averaged density matrix as \cite{Sekine-PRB}
\begin{eqnarray}\label{A-QLE}
\frac{\partial f }{{\rm d} t} +[H,f] + J(f)=0,
\end{eqnarray}
where $J(f)$ is the scattering term with impurities. $J(f)$ will be calculated in section~\ref{Gradient-J}. Taking the inhomogenous magnetization into account in the Hamiltonian, a rigorous definition of the distribution function follows from the one-particle density matrix $f_{{\bm r}_1{\bm r}_2}(t)$. The aim of this subsection is to derive the equation of motion for a semiclassical analog of the quantum mechanical density matrix, which is provided by the Wigner representation. The Wigner representation for the density matrix can be expressed as
\begin{eqnarray}\label{Wigner-DM-M}
f_{{\bm k},{\bm r}}=\int \frac{{\rm d}{\bm q}}{(2\pi)^d}{\rm e}^{i{\bm q}\cdot{\bm r}}f_{{\bm k}_+{\bm k}_-},
\end{eqnarray}
where $f_{{\bm k}_1{\bm k}_2}=f_{{\bm k}_+{\bm k}_-}\equiv \langle {\bm k}+{\bm q}/2|f|{\bm k}-{\bm q}/2\rangle$ with ${\bm k}=({\bm k}_1+{\bm k}_2)/2$ and ${\bm q}={\bm k}_1-{\bm k}_2$. 
The inverse transformation of the density matrix is
\begin{eqnarray}\label{Wigner-DM-I}
f_{{\bm k},{\bm r}}=\int\frac{ {\rm d}{\bm s}}{(2\pi)^d}{\rm e}^{-i{\bm k}\cdot{\bm s}}f_{{\bm r}_+{\bm r}_-},
\end{eqnarray}
where $f_{{\bm r}_1{\bm r}_2}=f_{{\bm r}_+{\bm r}_-}\equiv \langle {\bm r}+{\bm s}/2|f|{\bm r}-{\bm s}/2\rangle$ with ${\bm r}=({\bm r}_1+{\bm r}_2)/2$ and ${\bm s}={\bm r}_1-{\bm r}_2$. $d$ is the dimension of the system. Using equations~(\ref{Wigner-DM-M}) and (\ref{Wigner-DM-I}), we  transform the equation~(\ref{A-QLE}) into the quantum kinetic equation that captures the physics of an inhomogeneous magnetization  
\begin{eqnarray}\label{Kin}
\fl\pd{f_{{\bm k},{\bm r}}}{t}+ \frac{{\rm i}}{\hbar}[H_{\rm M}({\bm r}),f_{{\bm k},{\bm r}}] + \frac{1}{2\hbar}\bigg\{\pd{h_{\bm k}}{{\bm k}} \nabla f_{{\bm k},{\bm r}}\bigg\} -\frac{1}{2\hbar}\bigg\{ \nabla H_{\rm M}({\bm r}) \pd{f_{{\bm k},{\bm r}}}{{\bm k}}\bigg\} +J(f_{{\bm k},{\bm r}})=\mathcal{D}_{\bm E},
\end{eqnarray} 
where $\{{\bm a}{\bm b}\}={\bm a}\cdot{\bm b}+{\bm b}\cdot{\bm a}$, and the driving term due to the external electric field is 
\begin{eqnarray}
\mathcal{D}_{\bm E}=\frac{e}{\hbar} {\bm E}\cdot \pd{f_{{\bm k},{\bm r}}}{{\bm k}}.
\end{eqnarray}
 
\subsection{The scattering integral with gradient corrections}\label{Gradient-J}

\begin{figure}[h]
\begin{center}
\includegraphics[trim=0cm 0cm 0cm 0cm, clip, width=0.9\columnwidth]{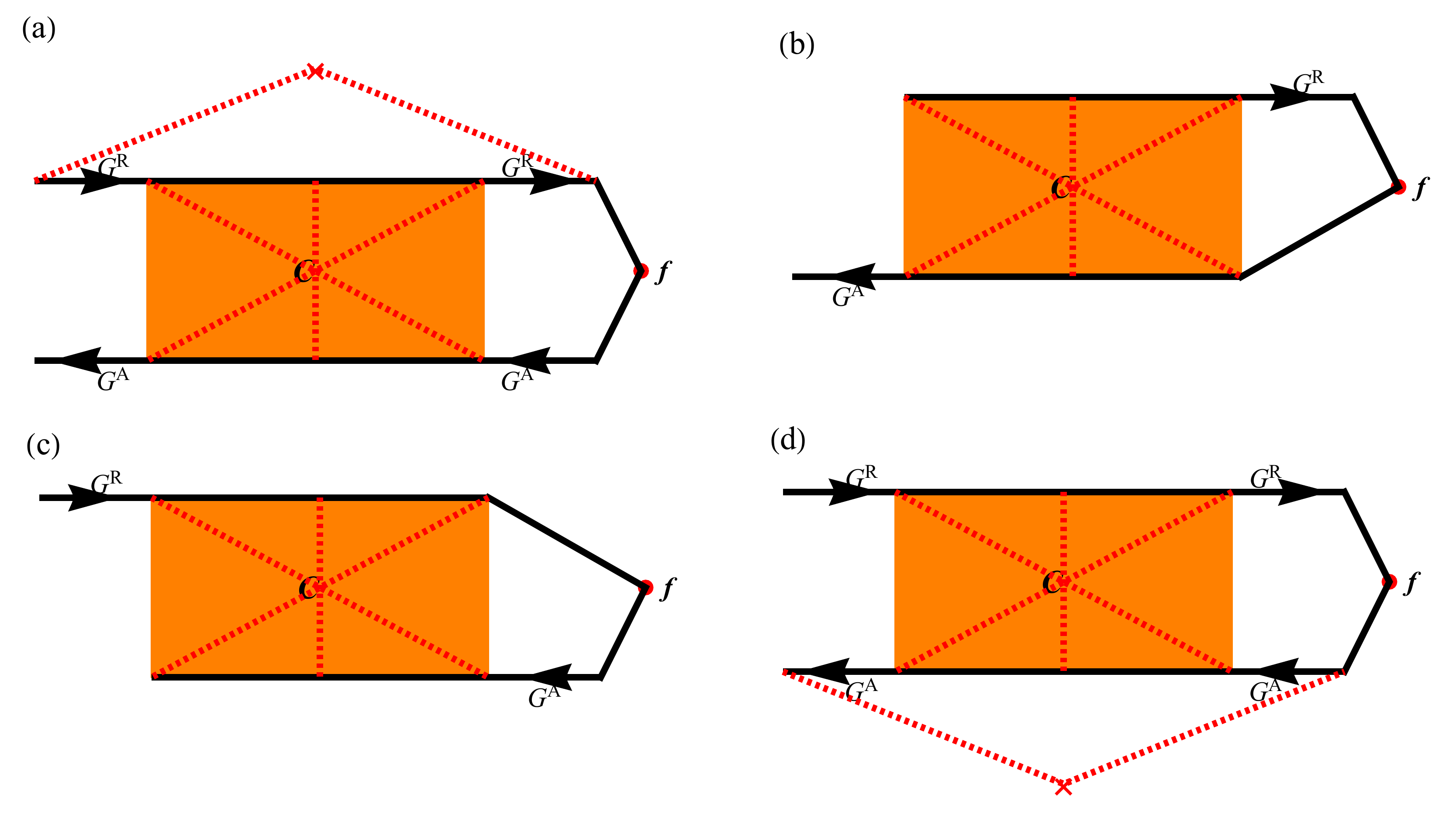}
\caption{\label{Fig1}The diagrams labeled by (a) (b) (c) (d) are corresponding to the scattering terms $J^{(1)}(f)$,$J^{(2)}(f)$,$J^{(3)}(f)$,$J^{(4)}(f)$ in equations~(\ref{1-term}),(\ref{2-term}),(\ref{3-term}) and (\ref{4-term}). The dashed red lines represent impurity lines. Solid right and left arrow lines represent Green's function $G^{\rm R}$ and $G^{\rm A}$, respectively. For simplicity, $G^{\rm R,A}(\epsilon)\equiv G^{\rm R,A}$. Red dot with two legs represents averaged density matrix $f$. The Cooperon operator $C$ is represented by orange rectangular box.}
\end{center}
\end{figure}

\begin{figure}[h]
\begin{center}
\includegraphics[trim=0cm 0cm 0cm 0cm, clip, width=1.0\columnwidth]{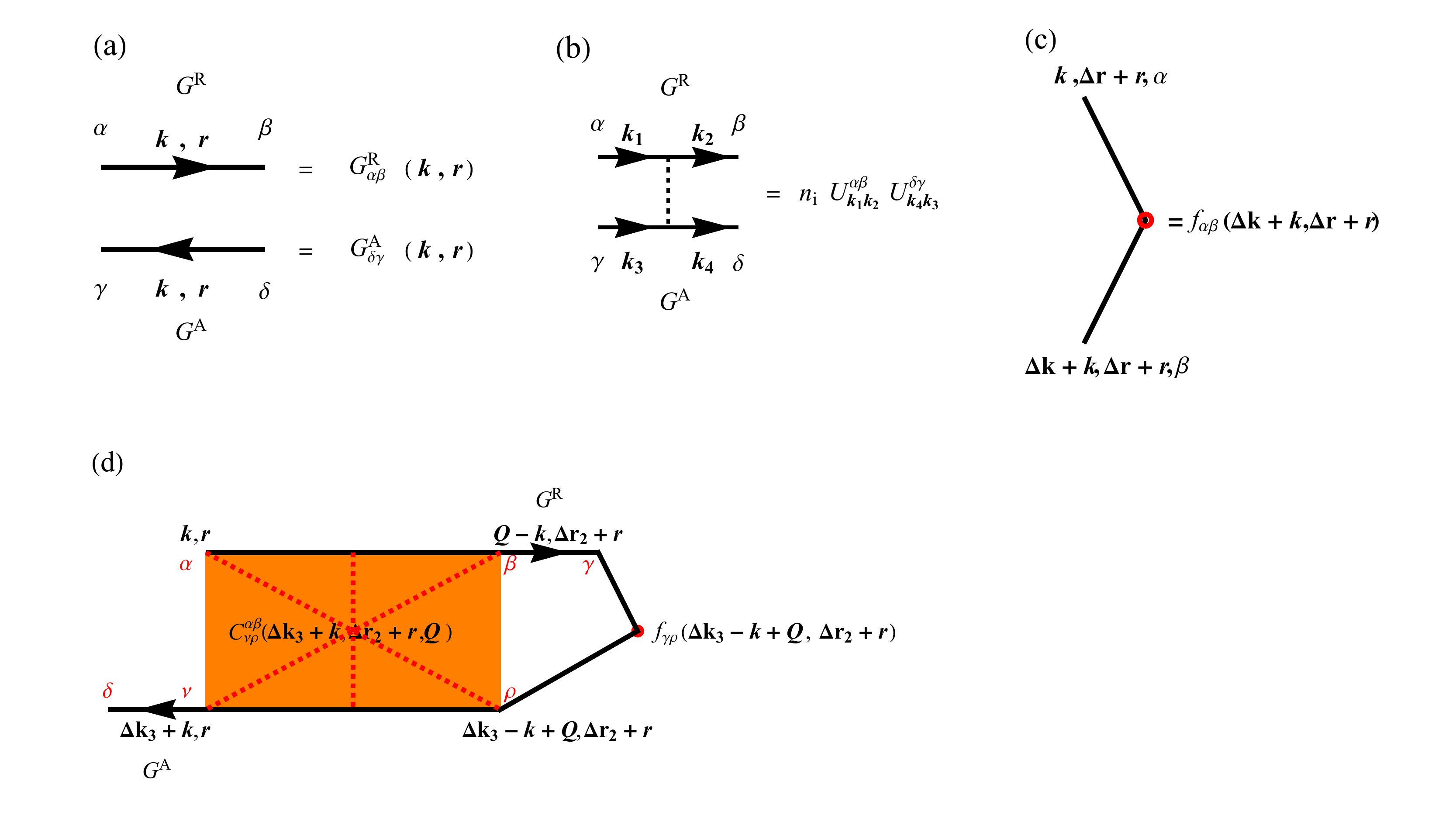}	
\caption{\label{Fig2} (a) Definition
of the Green’s functions as arrowed solid lines in which Greek letters are spin indices. (b)  Definition
of dashed lines: impurities lines expressed in both retarded and advanced cases. (c) Definition of the Wigner distribution function. (d) Indices label of $J^{(2)}(f_{{\bm k},{\bm r}})$ as an example to calculate the gradient correction in scattering term.}
\end{center}
\end{figure}

In section~\ref{Kin-Equation}, we obtain the general form of the kinetic equation for Wigner distribution function suitable for describing transport in inhomogeneous systems. To calculate the weak localization correction to transport we require the Wigner transformation of the scattering term $J(f)$ in equation~(\ref{A-QLE}) that yields $J(f_{{\bm k},{\bm r}})$. The operator form of scattering term can be expressed as 
\begin{eqnarray}
J(f)=\frac{1}{2\pi\hbar}\int^{\infty}_{\infty}{\rm d}\epsilon \langle [\hat{U},G^{\rm R}(\epsilon)[\hat{U},f 
]G^{\rm A}(\epsilon)],
\end{eqnarray}
which is corresponding to four elements as follows
\begin{eqnarray}
& J^{(1)}(f)=\frac{1}{2\pi\hbar}\int^{+\infty}_{-\infty} {\rm d}\epsilon \langle \hat{U}G^{\rm R}(\epsilon) \hat{U}f G^{\rm A}(\epsilon)\rangle \label{1-term},\\[3ex]
& J^{(2)}(f)=\frac{1}{2\pi\hbar}\int^{+\infty}_{-\infty}{\rm d}\epsilon \langle  \hat{U}G^{\rm R}(\epsilon)f  \hat{U}G^{\rm A}(\epsilon)\rangle, \label{2-term}\\[3ex]
& J^{(3)}(f)=\frac{1}{2\pi\hbar}\int^{+\infty}_{-\infty}{\rm d}\epsilon \langle G^{\rm R}(\epsilon) \hat{U}f G^{\rm A}(\epsilon) \hat{U}\rangle,\label{3-term}\\[3ex]
& J^{(4)}(f)=\frac{1}{2\pi\hbar}\int^{+\infty}_{-\infty}{\rm d}\epsilon \langle G^{\rm R}(\epsilon)f  \hat{U} G^{\rm A}(\epsilon) \hat{U}\rangle \label{4-term}.
\end{eqnarray}
$G^{\rm R,A}(\epsilon)$ are retarded and advanced Green's function. The $J^{(1)}(f)$ and $J^{(4)}(f)$ are complex conjugates and contribute to the dressed Cooperon in the weak localization. $J^{(2)}(f)$ and $J^{(3)}(f)$ are complex conjugate and contribute to the bare Cooperon in the weak localization. The diagrams corresponding to the four scattering terms are shown in figure~\ref{Fig1}. In the Born approximation, the disorder-averaged retarded Green’s functions in the eigenstate basis for a 2D electron gas with an inhomogeneous magnetization is
\begin{eqnarray}
G^{\rm R}_{\bm k}(\epsilon)=\left(
\begin{array}{cc}
 \frac{1}{\epsilon-\epsilon_{{\bm k},+}+\frac{{\rm i}\hbar}{2\tau}} &  0  \\
0  &    \frac{1}{\epsilon-\epsilon_{{\bm k},-}+\frac{{\rm i}\hbar}{2\tau}}  
\end{array}
\right).
\end{eqnarray}
When the spin polarization due to the interaction with the magnetization is taken into account, the lifetime $\tau$ becomes spin-dependent. For the most part of this paper, we neglect this spin-dependence, to avoid unnecessary complication, but we will restore it at the end. To determine the real space dependence of the Green's function, we express the Green's function in the Pauli spin eigenstate basis as $G^{\rm R}_{{\bm k},{\bm r}}(\epsilon )=T G^{\rm R}_{\bm k}(\epsilon)T^\dag$ with
 with \begin{eqnarray}
T=\left(
\begin{array}{cc}
 \cos \frac{\theta}{2}{\rm e}^{-\frac{{\rm i}\phi_r}{2}} & - \sin\frac{\theta}{2} {\rm e}^{-\frac{{\rm i}\phi_r}{2}}   \\
 \sin\frac{\theta}{2}{\rm e}^{\frac{{\rm i}\phi_r}{2}} &   \cos \frac{\theta}{2}{\rm e}^{\frac{{\rm i}\phi_r}{2}} 
\end{array}
\right).
\end{eqnarray}
For the short-range scalar impurity, the four constituents of the Born-approximation scattering term in the Wigner representation are 
\begin{eqnarray}\label{WT-initial}
\fl J^{(1)}(f_{{\bm k},{\bm r}})=\frac{2^d\times 2^d}{2\pi\hbar}\int^{\infty}_{\infty}{\rm d}\epsilon\int \frac{{\rm d}{\bm k}_1}{(2\pi)^d}\int {\rm d}\Delta {\bm r}_2\int {\rm d}\Delta{\bm r}_3\int \frac{{\rm d}\Delta{\bm k}_2}{(2\pi)^d}\int \frac{{\rm d}\Delta{\bm k}_3}{(2\pi)^d}{\rm e}^{-2{\rm i}\Delta{\bm k}_2\cdot\Delta{\bm r}_3}\\[3ex]
\times {\rm e}^{2{\rm i}\Delta{\bm k}_3\cdot\Delta{\bm r}_2}\langle U_{{\bm k}{\bm k}_1}G^{\rm R}_{{\bm k}_1,{\bm r}+\Delta{\bm r}_2-\Delta{\bm r}_3}(\epsilon)U_{{\bm k}_1{\bm k}}f_{{\bm k}+\Delta{\bm k}_2,{\bm r}+\Delta{\bm r}_2}G^{\rm A}_{{\bm k}+\Delta{\bm k}_3,{\bm r}+\Delta{\bm r}_3}(\epsilon)\rangle,\\[3ex]
\fl J^{(2)}(f_{{\bm k},{\bm r}})=\frac{2^d}{2\pi \hbar}\int^\infty_\infty {\rm d}\epsilon \int \frac{{\rm d} {\bm k}_1}{(2\pi)^d}\int {\rm d}\Delta {\bm r}_2 \int \frac{{\rm d}\Delta {\bm k}_3}{(2\pi)^d}{\rm e}^{2{\rm i}\Delta {\bm k}_3\cdot\Delta {\bm r}_2}\langle U_{{\bm k}{\bm k}_1} G^{\rm R}_{{\bm k}_1,{\bm r}+\Delta {\bm r}_2}(\epsilon) \\[3ex]
\times f_{{\bm k}_1+\Delta {\bm k}_3,{\bm r}+\Delta {\bm r}_2}U_{{\bm k}_1{\bm k}}G^{\rm A}_{{\bm k}+\Delta {\bm k}_3,{\bm r}}(\epsilon)\rangle,\\[3ex]
\fl J^{(3)}(f_{{\bm k},{\bm r}})=\frac{2^d}{2\pi \hbar}\int^\infty_\infty {\rm d}\epsilon \int \frac{{\rm d} {\bm k}_3}{(2\pi)^d}\int {\rm d}\Delta {\bm r}_2 \int \frac{{\rm d}\Delta {\bm k}_1}{(2\pi)^d}{\rm e}^{-2{\rm i}\Delta {\bm k}_1\cdot\Delta {\bm r}_2}\langle G^{\rm R}_{{\bm k}+\Delta {\bm k}_1,{\bm r}}U_{{\bm k}{\bm k}_3}(\epsilon)\\[3ex]
\times f_{{\bm k}_3+\Delta{\bm k}_1,{\bm r}+\Delta {\bm r}_2}G^{\rm A}_{{\bm k}_3,{\bm r}+\Delta{\bm r}_2}(\epsilon)U_{{\bm k}_3{\bm k}}\rangle,\\[3ex]
\fl J^{(4)}(f_{{\bm k},{\bm r}})=\frac{2^d\times 2^d}{2\pi\hbar}\int^{\infty}_{\infty}{\rm d}\epsilon\int \frac{{\rm d}{\bm k}_3}{(2\pi)^d}\int {\rm d}\Delta {\bm r}_1\int {\rm d}\Delta{\bm r}_2\int \frac{{\rm d}\Delta{\bm k}_1}{(2\pi)^d}\int \frac{{\rm d}\Delta{\bm k}_2}{(2\pi)^d}{\rm e}^{-2{\rm i}\Delta{\bm k}_1\cdot\Delta{\bm r}_2}\\[3ex]
\times {\rm e}^{2{\rm i}\Delta{\bm k}_2\cdot\Delta{\bm r}_1}\langle G^{\rm R}_{{\bm k}+\Delta{\bm k}_1,{\bm r}+\Delta{\bm r}_1}(\epsilon)f_{{\bm k}+\Delta{\bm k}_2,{\bm r}+\Delta{\bm r}_2}U_{{\bm k}{\bm k}_3}G^{\rm A}_{{\bm k}_3,{\bm r}+\Delta{\bm r}_2-\Delta{\bm r}_1}(\epsilon)U_{{\bm k}_3{\bm k}}\rangle.
\end{eqnarray}
The $2^d$ comes from Jacobian transformation in the integral. The density matrix and Green's function in equations above are $2\times2$ matrices. The gradient correction to the scattering then can be calculated by expanding the Green's function and density matrix in small quantities. Taking $J^{(2)}(f_{{\bm k},{\bm r}})$ as an example [see figure~\ref{Fig2}(d)],
\begin{eqnarray}
G^{\rm R}_{{\bm k}_1,{\bm r}+\Delta {\bm r}_2}(\epsilon) \approx G^{\rm R}_{{\bm k}_1,{\bm r}}(\epsilon)+\Delta {\bm r}_2 \pd{G^{\rm R}_{{\bm k}_1,{\bm r}}(\epsilon)}{\bm r},\\[3ex]
f_{{\bm k}_1+\Delta {\bm k}_3,{\bm r}+\Delta {\bm r}_2} \approx f_{{\bm k}_1,{\bm r}}+\Delta {\bm r}_2\pd{f_{{\bm r},{\bm k}_1} }{\bm r}+\Delta {\bm k}_3 \pd{f_{{\bm k}_1,{\bm r}}}{{\bm k}_1},\\[3ex]
G^{\rm A}_{{\bm k}+\Delta {\bm k}_3,{\bm r}}(\epsilon)\approx G^{\rm A}_{{\bm k},{\bm r}}(\epsilon)+\Delta {\bm k}_3\pd{G^{\rm A}_{{\bm k},{\bm r}}(\epsilon)}{\bm k},\\[3ex]
 C(\Delta{\bm k}_3\!+\!{\bm k},\Delta{\bm r}_2\!+\!{\bm r},{\bm Q})=C({\bm Q},{\bm r})\!+\!\Delta{\bm k}_3\pd{C({\bm Q},{\bm r} ) }{\bm k} +\Delta{\bm r}_2 \pd{C({\bm Q},{\bm r})}{\bm r}.
\end{eqnarray}
Then we can integrate over the small quantities using
\begin{eqnarray}
\int {\rm d}\Delta{\bm r}_2 \int  \frac{{\rm d}\Delta {\bm k}_3}{(2\pi)^d}{\rm e}^{2{\rm i}\Delta {\bm k}_3\cdot\Delta {\bm r}_2}\Delta{\bm r}_2\Delta{\bm k}_3=-\frac{1}{2{\rm i}}\frac{1}{2^d}.
\end{eqnarray}
At this stage we are ready to find the gradient correction to the Born-approximation scattering term as well as the maximally crossed diagrams. In the Pauli basis
\begin{eqnarray}
\fl [J^{0,(2)}_{\rm DB}(f_{{\bm k},{\bm r}})]^{\alpha\delta}=\frac{1}{2\pi\hbar}\sum_{\beta\gamma\rho\nu}\int {\rm d}\epsilon  \int \frac{{\rm d}{\bm k}_1}{(2\pi)^d}\langle U^{{\bm k}{\bm k}_1}_{\alpha \beta}G^{\rm R}_{\beta\gamma}({\bm k}_1,{\bm r})f_{\gamma \rho}({\bm k}_1,{\bm r})U^{{\bm k}_1{\bm k}}_{\rho\nu}G^{\rm A}_{\nu\delta}({\bm k},{\bm r})\rangle,\\[3ex]
\fl [J^{{\rm g},(2)}_{\rm DB}(f_{{\bm k},{\bm r}})]^{\alpha\delta}=-\frac{1}{2{\rm i}}\frac{1}{2\pi\hbar}\sum_{\beta\gamma\rho\nu}\int {\rm d}\epsilon  \int \frac{{\rm d}{\bm k}_1}{(2\pi)^d}\Big[\langle U^{{\bm k}{\bm k}_1}_{\alpha \beta} \pd{G^{\rm R}_{\beta\gamma}({\bm k}_1,{\bm r})f_{\gamma \rho}({\bm k}_1,{\bm r})}{\bm r}\\[3ex]
\times U^{{\bm k}_1{\bm k}}_{\rho\nu}\pd{G^{\rm A}_{\nu\delta}({\bm k},{\bm r})}{\bm k}\rangle+\langle U^{{\bm k}{\bm k}_1}_{\alpha \beta}\pd{G^{\rm R}_{\beta\gamma}({\bm k}_1,{\bm r})}{\bm r}\pd{f_{\gamma \rho}({\bm k}_1,{\bm r})}{{\bm k}_1}U^{{\bm k}_1{\bm k}}_{\rho\nu}G^{\rm A}_{\nu\delta}({\bm k},{\bm r})\rangle\Big],\\[3ex]
\fl [J^{0,(2)}_{\rm WL}(f_{{\bm k},{\bm r}})]^{\alpha\delta}=\sum_{\beta\gamma\rho\nu}\frac{1}{2\pi \hbar}\int\!\int \frac{{\rm d}\epsilon{\rm d}{\bm Q}}{(2\pi)^d}G^{\rm R}_{\beta\gamma}({\bm Q}\!-\!{\bm k},{\bm r})C^{\alpha\beta}_{\nu\rho}({\bm Q},{\bm r}) f_{\gamma\rho}({\bm Q}\!-\!{\bm k},{\bm r}) 
G^{\rm A}_{\nu\delta}({\bm k},{\bm r}),\\[3ex]
\fl [J^{{\rm g},(2)}_{\rm WL}(f_{{\bm k},{\bm r}})]^{\alpha\delta}=-\frac{1}{2{\rm i}}\sum_{\beta\gamma\rho\nu}\frac{1}{2\pi \hbar}\int {\rm d}\epsilon \int \frac{{\rm d}{\bm Q}}{(2\pi)^d}\\[3ex]
\fl \times \Big[\pd {G^{\rm R}_{\beta\gamma}({\bm Q}\!-\!{\bm k},{\bm r})}{{\bm r}} \big[\pd{C^{\alpha\beta}_{\nu\rho}({\bm Q},{\bm r})}{\bm Q}f_{\gamma\rho}({\bm Q}\!-\!{\bm k},{\bm r})+C^{\alpha\beta}_{\nu\rho}({\bm Q},{\bm r}) \pd{f_{\gamma\rho}({\bm Q}\!-\!{\bm k},{\bm r}) }{({\bm Q}\!-\!{\bm k})}\big]G^{A}_{\nu\delta}({\bm k},{\bm r})\\[3ex]
\fl +\pd{}{\bm r}\big[C^{\alpha\beta}_{\nu\rho}({\bm Q},{\bm r})G^{\rm R}{\beta\gamma}({\bm Q}-{\bm k},{\bm r}) f_{\gamma\rho}({\bm Q}\!-\!{\bm k},{\bm r})\big]\pd{G^{\rm A}_{\nu\delta}({\bm k},{\bm r})}{{\bm k}} \\[3ex]
\fl +\big[\pd{C^{\alpha\beta}_{\nu\rho}({\bm Q},{\bm r})}{\bm r}\pd{f_{\gamma\rho}({\bm Q}\!-\!{\bm k},{\bm r})}{({\bm Q}\!-\!{\bm k})}\!+\!\pd{C^{\alpha\beta}_{\nu\rho}({\bm Q},{\bm r})}{\bm Q}\pd{f_{\gamma\rho}({\bm Q}\!-\!{\bm k},{\bm r})}{\bm r}\big]G^{\rm R}_{\beta\gamma}({\bm Q}\!-\!{\bm k},{\bm r}) G^{A}_{\nu\delta}({\bm k},{\bm r})\Big].
\end{eqnarray}
Here ${\bm Q}$ is the momentum change after a multiple scattering process with disorder. $C({\bm Q},{\bm r})$ is the Cooperon, which is dependence both on momenta ${\bm Q}$ and real space location ${\bm r}$. Summing over all the terms in four scattering elements, we get \\x$J(f_{{\bm k},{\bm r}})=J^0_{\rm DB}(f_{{\bm k},{\bm r}})+J^0_{\rm WL}(f_{{\bm k},{\bm r}})+J^{\rm g}_{\rm DB}(f_{{\bm k},{\bm r}})+J^{\rm g}_{\rm WL}(f_{{\bm k},{\bm r}})$.
$J^0_{\rm DB}(f_{{\bm k},{\bm r}})$ and $J^0_{\rm WL}(f_{{\bm k},{\bm r}})$ are is the Born-approximation and  weak localization correction to the scattering term without gradient correction, respectively. $J^{\rm g}_{\rm DB}(f_{{\bm k},{\bm r}})$ and $J^{\rm g}_{\rm WL}(f_{{\bm k},{\bm r}})$ are the gradient correction to the Born-approximation scattering and to the weak localization scattering term, respectively. 

\subsection{Recovering the topological Hall effect}\label{THE}
In this section we solve Eq.~(\ref{Kin}) and demonstrate that the original topological Hall effect formula of Ref.~\cite{Bruno-THE-PRL-2004} can be reproduced by including the scattering term $J^0_{\rm DB}(f_{{\bm k},{\bm r}})$. The calculation of the topological Hall effect in the density matrix formalism is decomposed into part for the charge density part and a part for the spin density. The solution of equation~(\ref{Kin}) in equilibrium is  separated into $f_{0}({\bm k},{\bm r})=n_{0}({\bm k})+{S}_{0}({\bm k},{\bm r})$ with $n_{0}({\bm k})=\frac{1}{2}( f_{0,+}+ f_{0,-})$ the charge part and ${S}_{0}({ \bm k},{\bm r})=\frac{1}{2}( f_{0,+}- f_{0,-}){\bm \sigma}\cdot{\bm n}({\bm r})$ the spin-part, while $f_{0,\pm}=f_0(\epsilon_{k}\pm M)$ is the Fermi-Dirac distribution function for the conduction band and valence band resepctively. The spin direction can be mapped within the frame ${\bm n}({\bm r}),  \partial_i {\bm n}({\bm r}), \partial_i {\bm n}({\bm r})\times  {\bm n}({\bm r})$, where $\sigma_{\parallel}={\bm \sigma}\cdot{\bm n}({\bm r})$ and $\sigma_{\perp}={\bm \sigma}\cdot \partial_i {\bm n}({\bm r})+ {\bm \sigma}\cdot \partial_i {\bm n}({\bm r})\times  {\bm n}({\bm r})$. Then we can decompose the spin density matrix into parallel and perpendicular parts with ${S}({\bm k},{\bm r})=S_{\parallel}({\bm k},{\bm r})+S_{\perp}({\bm k},{\bm r})$. In matrix language, $S_{\parallel}({\bm k},{\bm r})$ is parallel to the Hamiltonian $H_{\rm M}({\bm r})$ i.e. it commutes with the Hamiltonian, while the remainder $S_{\perp}({\bm k},{\bm r})$ is orthogonal to  the Hamiltonian $H_{\rm M}({\bm r})$. Because $[H_{\rm M}({\bm r}),S_{\parallel}({\bm k},{\bm r})]=0$, the parallel part $S_{\parallel}({\bm k},{\bm r})$ does not change in time under the action of the time evolution operator $e^{iH_0t/\hbar}$. Therefore $S_{\parallel}({\bm k},{\bm r})$ represents the conserved spin fraction and $S_{\perp}({\bm k},{\bm r})$ is the precessing spin fraction.

To find the density matrix out of equilibrium, we consider an electric field ${\bm E}$ applied in the $\hat{\bm x}$ and obtain the driving term. We solve the kinetic equation in Pauli basis, the solution of the kinetic equation is $f_{\bm E}({\bm k},{\bm r}) =n_{\bm E}({\bm k},{\bm r}) + S_{\bm E}({\bm k},{\bm r})$ can be found using the blueprint of reference \cite{Sekine-PRB}.
\begin{eqnarray}\label{E-DM}
\ba
n_{\bm E}({\bm k},{\bm r}) =&\dps \frac{e{\bm E}\tau}{2\hbar}\cdot\pd{}{{\bm k}}( f_{0,+}+ f_{0,-}), \\[3ex]
S_{\bm E}({\bm k},{\bm r})=&\dps \frac{e{\bm E}\tau}{2\hbar}\cdot\pd{}{{\bm k}}\Big[( f_{0,+}- f_{0,-}){\bm \sigma}\cdot{\bm n}({\bm r}) \Big],
\ea
\end{eqnarray}
where $1/\tau=2\pi n_{\rm i}u^2\hbar/\rho(\epsilon_{\rm F})$ is the momentum relaxation time and $\epsilon_{\rm F}$ is the Fermi energy. $\rho(\epsilon_{\rm F})=m/(2\pi\hbar^2)$ is the density of state for two dimensional electron gas.   

To calculate the topological Hall effect, we regard the gradient terms in the quantum kinetic equation as perturbation, and fed $f_{\bm E}({\bm k},{\bm r})$ into the kinetic equation to get the effective driving term. 
\begin{eqnarray}
 \mathcal{D}^{(1)}_{\nabla}=-\frac{\hbar {\bm k}}{m}\cdot \nabla f_{\bm E}({\bm k},{\bm r})+\frac{1}{2\hbar}\Big\{\nabla H_{\rm M}({\bm r})\pd{f_{\bm E}({\bm k},{\bm r})}{\bm k}\Big\},
\end{eqnarray}
where we have $\mathcal{D}^{(1)}_{\nabla}\propto\hat{k}_i {\bm \sigma}\cdot \partial_i{\bm n}({\bm r})$ which contributes to the perpendicular part of density matrix $S_{\perp}({\bm k},{\bm r})$. Solving the kinetic equation as follows
\begin{eqnarray}\label{Kin-THE-1}
\pd{S^{(1)}_{\nabla,\perp}}{t}+ \frac{\rm i}{\hbar}[H_{\rm M}({\bm r}),S^{(1)}_{\nabla,\perp}] +J^0_{\rm DB}(S^{(1)}_{\nabla,\perp})=\mathcal{D}^{(1)}_{\nabla},
\end{eqnarray} 
we will get the perpendicular spin density matrix in first-order gradient written as $S^{(1)}_{\nabla,\perp}$, and the solution is
\begin{eqnarray}
\fl S^{(1)}_{\nabla,\perp}=\lim_{\eta\rightarrow 0}\int^{\infty}_0dt e^{-\eta t}e^{-iH_Mt/\hbar}\mathcal{D}^{(1)}_{\nabla}e^{iH_Mt/\hbar}=\frac{\hbar}{2}\hat{k}_i{\bm \sigma}\cdot[\partial_i{\bm n}({\bm r})\times{\bm n}({\bm r})]\frac{e{\bm E}\cdot\hat{\bm k}\tau}{\hbar^2k}\pd{f_{0}}{k}.
\end{eqnarray}
Similarly, the driving term for second-order gradient can be expressed as
\begin{eqnarray}
\mathcal{D}^{(2)}_{\nabla}= -\frac{\hbar {\bm k}}{m}\cdot \nabla S^{(1)}_{\nabla,\perp}+\frac{1}{2\hbar}\Big\{\nabla H_{\rm M}({\bm r}) \pd{S^{(1)}_{\nabla,\perp}}{\bm k}\Big\}.
\end{eqnarray}
With 
\begin{eqnarray}
\fl \hat{\bm k}\cdot \nabla\Big[{\bm \sigma}\cdot[\partial_i{\bm n}({\bm r})\times{\bm n}({\bm r})] \Big]\approx \hat{k}_j{\bm \sigma}\cdot [\partial_i{\bm n}({\bm r})\times \partial_j{\bm n}({\bm r})],\\[3ex]
\fl \Big\{\hat{\bm k}\cdot \nabla[{\bm \sigma}\cdot {\bm n}({\bm r})]\Big[{\bm \sigma}\cdot[\partial_i{\bm n}({\bm r})\times{\bm n}({\bm r})]\Big]\Big\}= 2{\bm n}({\bm r})\cdot[\partial_i {\bm n}({\bm r})\times \partial_j {\bm n}({\bm r})],
\end{eqnarray}
we have $\mathcal{D}^{(2)}_{\nabla}$ contributes to the charge density matrix and parallel part of spin density matrix. Solving $J^0_{\rm DB}(f^{(2)}_\nabla)=\mathcal{D}^{(2)}_{\nabla}$, we get $f^{(2)}_\nabla=n^{(2)}_{\nabla}+S^{(2)}_{\nabla,\parallel}$.
\begin{eqnarray}
n^{(2)}_{\nabla}=M\hat{k}_i\hat{k}_j\pd{}{k}\Big[ \frac{e{\bm E}\cdot\hat{\bm k}\tau^2}{\hbar^2k}\Big]{\bm n}({\bm r})\cdot[\partial_i {\bm n}({\bm r})\times \partial_j {\bm n}({\bm r})],
\end{eqnarray}
 \begin{eqnarray}
S^{(2)}_{\nabla,\parallel}
 =\hat{k}_i\hat{k}_j{\bm \sigma}\cdot[\partial_i{\bm n}({\bm r})\times\partial_j{\bm n}({\bm r})] \Big[\frac{e{\bm E}\cdot\hat{\bm k}\tau^2}{2m}\pd{f_{0}}{k}\Big].
\end{eqnarray}
The topological Hall current can be written as $j^y=
\Tr[\hat{v}_y n^{(2)}_{\nabla}]$, where $\hat{v}_y$ is the velocity operator in the transverse direction. Then we have the topological Hall conductivity as
\begin{eqnarray}\label{THE-conductivity}
\ba
&\dps \sigma^{\rm T}_{yx}= - \frac{e^2}{h}\frac{e\tau}{m} \frac{M\tau}{\hbar} \frac{h}{4 \pi e}{\bm n}({\bm r})\cdot[\partial_i{\bm n}({\bm r})\times \partial_j{\bm n}({\bm r})],
\ea
\end{eqnarray}
 where $B_t=\frac{\phi_0}{4 \pi}{\bm n}({\bm r})\cdot \big[\partial_i{\bm n}({\bm r})\times \partial_j{\bm n}({\bm r})\big]$ and $\phi_0=\frac{hc}{e}$ is the magnetic flux quantum. The topological Hall conductivity obtained here is in exact agreement with reference \cite{Bruno-THE-PRL-2004}, where the total $\sigma^{\rm T}_{yx}=\sum_s\sigma^s_{yx}$ with $s=\uparrow, \downarrow$.

\subsection{Weak localization correction to the longitudinal conductivity}\label{WLxx}

The weak localization correction to the longitudinal current can be calculated by including $J^0_{\rm WL}(f_{{\bm k},{\bm r}})$. In this paper, we consider scalar disorder scattering, therefore the impurity line in Fig.~\ref{Fig2}(b) is reduced to $n_{\rm i}u^2$. The calculation of Cooperon is standard. For system with strong exchange interaction, the twisted Cooperon $\tilde{C}(Q)$ in the eigenstate basis takes the form
\begin{eqnarray}
\ba
&\dps \tilde{C}(Q)\approx\frac{\hbar}{2\pi \tau \rho(\epsilon_{\rm F})}\frac{2}{\tau^2 v^2_{\rm F}Q^2} \left(
\begin{array}{c c c c}
1 &  0 &  0 & 0 \\
0  & 0 &   0 & 0\\
0  &  0 &   0 & 0\\
 0 & 0   &   0 & 1
\end{array}
\right).
\ea
\end{eqnarray}
Solving the kinetic equation $
J^0_{\rm DB}(f^{(0)}_{\rm WL})=-J^0_{\rm WL}[f_{\bm E}({\bm k},{\bm r})]$, we find the correction to the density matrix
\begin{eqnarray}
f^{(0)}_{\rm WL}=-\frac{2}{h}\sum_{\bm Q}  \frac{1}{\tau v^2_{\rm F}Q^2 \rho(\epsilon_{\rm F})} f_{\bm E}({\bm k},{\bm r}).
\end{eqnarray}
The weak localization correction to the longitudinal current is calculated as $\delta j^x=\Tr[\hat{v}_xf^{(0)}_{\rm WL}]$. This yields the standard result $\delta \sigma_{xx} = - \frac{e^2}{h\pi} \ln \frac{l_\phi}{l}$, where $l_\phi, l$ are the phase coherence length and mean free path respectively. 

\subsection{Weak localization correction to the topological Hall effect}\label{WL-THE}

In sections~\ref{THE} and~\ref{WLxx}, we have solved the density matrix pertubatively without the gradient scattering term. In this section, we calculate the density matrix with the full $J(f_{{\bm k},{\bm r}})$. The usual weak localization term in the Drude conductivity does not enter the gradient correction due to an inhomogenous magnetization, while $J^{\rm g}_{\rm DB}(f_{{\bm k},{\bm r}})$ and $J^{\rm g}_{\rm WL}(f_{{\bm k},{\bm r}})$ must be included. The kinetic equation with an additional gradient scattering term will give rise to a weak localization correction of topological Hall effect. Solving the kinetic equation to determine the density matrix $f_{\nabla,{\rm WL}}$ with both gradient and weak localization corrections is key to our approach. We write the kinetic equation explicitly as follows
\begin{eqnarray}
\fl \pd{f_{\nabla,{\rm WL}}}{t}+ \frac{\rm i}{\hbar}[H_{\rm M}({\bm r}),f_{\nabla,{\rm WL}}] +J^0_{\rm DB}(f_{\nabla,{\rm WL}})\\[3ex]
\fl =-\frac{\hbar {\bm k}}{m}\cdot \nabla f^{(0)}_{\rm WL}+\frac{1}{2\hbar}\Big\{\nabla H_{\rm M}({\bm r}) \pd{f^{(0)}_{\rm WL}}{\bm k}\Big\}-J^0_{\rm WL} (f^{(2)}_{\nabla})- J^{\rm g}_{\rm DB}(f^{(0)}_{\rm WL})-J^{\rm g}_{\rm WL}[f_{\bm E}({\bm k},{\bm r})],
\end{eqnarray} 
We found that the contributions from $-J^0_{\rm WL} (f^{(2)}_{\nabla})$ and $-\frac{\hbar {\bm k}}{m}\cdot \nabla f^{(0)}_{\rm WL}+\frac{1}{2\hbar}\Big\{\nabla H_{\rm M}({\bm r}) \pd{f^{(0)}_{\rm WL}}{\bm k}\Big\}$ cancel each other. The contribution from  $J^{\rm g}_{\rm DB}(f^{(0)}_{\rm WL})=0$ in the system we consider here. To calculate the contribution from $J^{\rm g}_{\rm WL}[f_{\bm E}({\bm k},{\bm r})]$, we need to transform the twisted Cooperon $\tilde{C}({\bm Q})$ to the Pauli basis using $\tilde{C}({\bm Q},{\bm r})=T\otimes T\tilde{C}({\bm Q}) T^\dag\otimes T^\dag $ and exchange the spin indices to yield the Cooperon $C({\bm Q},{\bm r})$. Feeding the Cooperon matrix element $C^{\alpha\beta}_{\nu\rho}({\bm Q},{\bm r})$, the density matrix element $f_{\bm E}({\bm k},{\bm r})$ as well as the Green's function matrix element $G^{\rm R,A}_{\alpha\beta}({\bm k},{\bm r})$ into $J^{\rm g}_{\rm WL}[f_{\bm E}({\bm k},{\bm r})]$, we obtain the density matrix contribution to first order in the gradient $f^{(1)}_{\nabla,{\rm WL}}$ as the solution of $
J^0_{\rm DB}(f^{(1)}_{\nabla,{\rm WL}})=-J^{\rm g}_{\rm WL}[f_{\bm E}({\bm k},{\bm r})]$. To find the density matrix $f^{(2)}_{\nabla,{\rm WL}}$ to second order in the gradient, we go through the same process as that in the topological Hall effect calculation with 
\begin{eqnarray}
\mathcal{D}^{(2)}_{\nabla,{\rm WL}}= -\frac{\hbar {\bm k}}{m}\cdot \nabla f^{(1)}_{\nabla,{\rm WL}}+\frac{1}{2\hbar}\Big\{\nabla H_{\rm M}({\bm r}) \pd{f^{(1)}_{\nabla,{\rm WL}}}{\bm k}\Big\}.
\end{eqnarray}
Finally, we get the density matrix contributing to the charge current in the transverse direction 
\begin{eqnarray}
\fl \qquad n^{(2)}_{\nabla,{\rm WL}}=\sum_{\bm Q}\frac{4\pi}{M}\frac{1}{\rho(\epsilon_{\rm F})}\frac{1}{\tau^2v^2_{\rm F}Q^2}\frac{\partial^2 n_{\bm E}({\bm k},{\bm r})}{\partial k^2}\hat{k}_i\hat{k}_j{\bm n}({\bm r})\cdot[\partial_i{\bm n}({\bm r})\times\partial_j{\bm n}({\bm r})].
\end{eqnarray}
After taking trace $\delta j^y=\Tr[\hat{v}_y n^{(2)}_{\nabla,{\rm WL}}]$, we get the gradient correction to the topological Hall conductivity 
\begin{eqnarray}\label{Final-WAL}
\delta\sigma^{\rm g}_{yx}= \frac{e^2}{\hbar} \frac{1}{4\pi^2 } \frac{\hbar }{\epsilon_{\rm F}\tau}\frac{\hbar}{M\tau}\frac{e\tau B_t}{m}\ln \frac{l_{\phi}}{l}.
\end{eqnarray}

\section{Results and Discussion}\label{Result} 

The primary result of the paper is the weak localization correction in the absence of an external magnetic field, which is equation~(\ref{Final-WAL}). The averaged component of the topological magnetic field perpendicular to the 2D plane is given by $B_t\equiv Q\phi_0/\mathcal{A}$, where $Q$ is an integer number and $\mathcal{A}$ is an area of the chiral spin texture which is also proportional to the density of isolated chiral spin texture.
The ration between correction to topological Hall resistivity and topological Hall resistivity is 
\begin{eqnarray}
\frac{\delta\rho^{\rm g}_{yx}}{\rho^{\rm T}_{yx}}=-\frac{1}{2\pi} \frac{\hbar}{\epsilon _{\rm F}\tau}\bigg(\frac{\hbar}{M\tau}\bigg)^2\ln\frac{l_\phi}{l},
\end{eqnarray}
which is around $10^{-3}$ order of topological Hall resistivity for systems such as dilute magnetic semiconductor. For our theory to be applicable, we require $\hbar/\epsilon_{\rm F}\tau <1$. The $<$ sign means the Fermi wave vector is greater than the inverse of the mean free path, but not much greater, so that disorder effects are observable. Conventionally, when one restricts one’s attention to the weak momentum scattering regime where $\hbar/\epsilon_{\rm F}\tau \ll1$, yet, in this case, weak localization/antilocalization corrections tend to be negligible. In this paper, we consider the strong exchange interaction case with adiabatic parameter $M\tau/\hbar>1$. 

All electrical measurements of magnetic skyrmions have been based on the interpretation that the topological Hall resistivity
is proportional to the number of skyrmions multiplied by the
magnetic flux quantum \cite{Bruno-THE-PRL-2004}. Significant progress has been achieved due to recent advances in imaging techniques that have led to the discovery of different material platforms hosting skyrmions with size ranging from sub 100 nm \cite{Tr-Fe-Co-Pt-THE,Skymions-NanoTech,Skymions-NL-100nm,Skymions-Tr-Fe-Co-Pt} down to sub 10 nm \cite{PhysRevLett.114.177203,Skymions-sub10nm}. These sizes correspond to emergent magnetic fields ranging from 1 to 4000 T. In the Berry phase picture, spin of conduction electrons must fully align to the local magnetic moment during their movement, an adiabatic process is needed, which is only satisfied for strong coupling. So far, most experimental works on the topological Hall effect belong to this case. In weak coupling case, especially in the diffusive regime, the condition of adiabatic picture fails, and nonadiabatic contribution needs to be taken into account. Theory predictions of nonadiabatic effect have been made \cite{THE-W-S-theory,THE-weak-coupling-PRB,WeakCoupling-THE-RPB-2019}, while experimental confirmation is still lack\cite{WANG-THE-review}. In this paper, our theory has been simplified and based on strong exchange interaction where the emergent topological field is an averaged component perpendicular to the 2D plane, and the dispersion of conduction electron is chosen to be quadratic. Therefore, our theory is applicable to most current topological Hall experiments. 

With this in mind, it is important to note the principal difference in the topological Hall effect behavior for dense and dilute skyrmion systems. The first case corresponds to the dense skyrmionic system with the distance between the skyrmions $d_{\rm sk}$ of the same order as their size $a$. The second case is a dilute system with skyrmions randomly distributed with large distances between each other, so that $d_{\rm sk}\ll a$. A consistent theory of the topological Hall effect has to take into account the evolution of a skyrmion arrangement upon changing the external conditions. Especially, the topological Hall resistivity is usually measured under the external magnetic field. The skymion size decreases with increasing external magnetic field. The skyrmion sheet density, in principle, can depend on the magnetic field thus leading to the transition from dense to dilute regime \cite{Skymions-NanoTech,PhysRevB.98.214407,Tr-Fe-Co-Pt-THE}.  Skyrmions size in ferromagnetic layer composition Ir/Fe/Co/Pt multi-layers and their sheet density can be varied over a wide range by changing the composition of the structure and by tuning the external parameters \cite{PhysRevB.98.214407,Skymions-Tr-Fe-Co-Pt}. Smallest skyrmion of few-nm size was detected in a strong magnetic field in a PdFe bilayer on Ir(111) surface using spin-polarized scanning tunneling microscopy. Fe/Ir(111)-PdFe/Ir(111) became the model systems for study of different regimes of magnetic ordering from single skyrmions to skyrmion lattices. The considered theory of weak localization correction to the topological Hall effect in a dilute array of chiral spin textures is valid when the mean free path $l$ is larger than skymions size $a$. We have considered the dilute regime, when the scattering rate on spin textures is much smaller than that on nonmagnetic impurities $\tau_s=\tau$. Neglecting thermal effects \cite{TNE-Skymions-PRL}, we assume that the scattering on nonmagnetic impurities gives the dominating contribution to the transport scattering time and determines the mean free path.  

For instance, scattering off nonmagnetic impurities gives the dominant contribution to the transport scattering time of nanoscale skyrmions observed in atomically thin magnetic layers, such as PdFe/Ir(111) \cite{Nanoscale-Skymions-dilute}. In thin ferromagnetic films PdFe/Ir, the conduction electrons in the magnetic film interact with skyrmions as well as with nonmagnetic impurities. The estimated parameters in the Tr/Fe/Co/Pt multilayers are $M = 0.6$ $eV$ and $\epsilon_{\rm F}=5$ $eV$. With effective in-plane mass $m=m_0$. If we estimate the momentum relaxation time $\tau=a/v_{\rm F}$ with $a=50$ nm, we have $M\tau/\hbar\approx30$. In addition, the large Fermi energy in the system limit the regime in weak momentum scattering $\hbar/\epsilon_{\rm F}\tau\ll 1$, therefore the weak localization correction to the topological Hall resistivity in this system is negligible.  

\begin{figure}[h]
\begin{center}
\includegraphics[trim=0cm 0cm 0cm 0cm, clip, width=0.45\columnwidth]{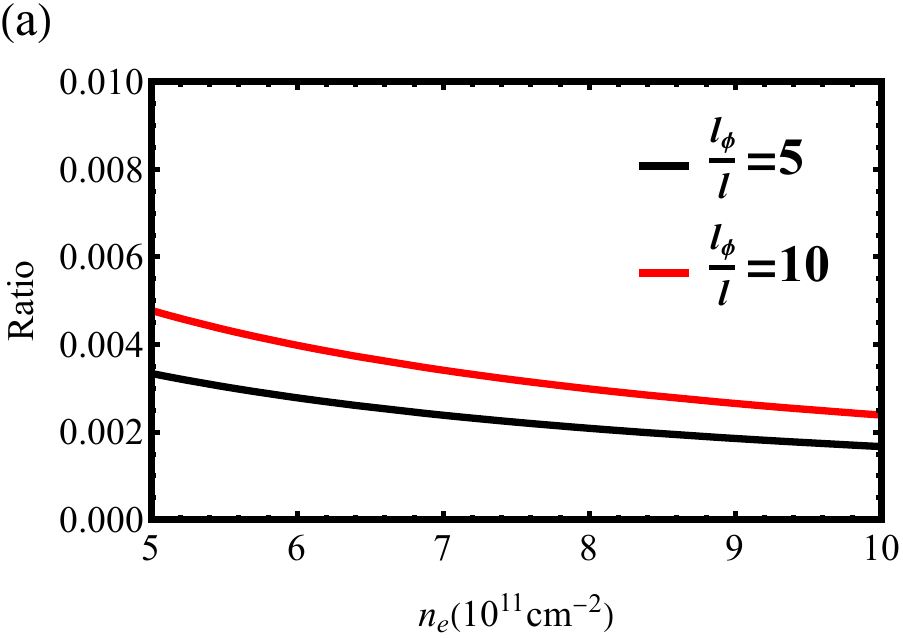}
\includegraphics[trim=0cm 0cm 0cm 0cm, clip, width=0.45\columnwidth]{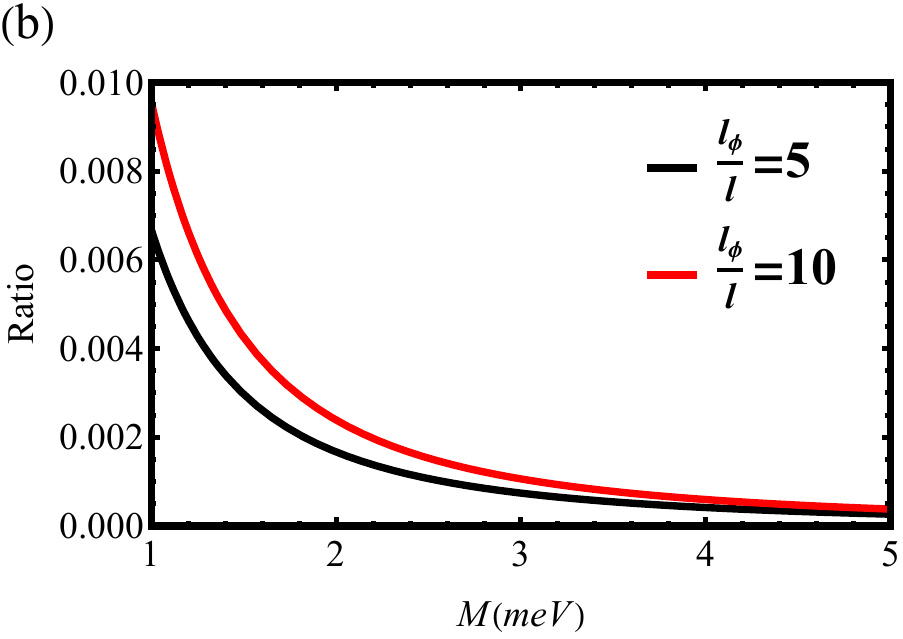}
		\caption{\label{Fig3}(a) The carrier density dependence of ${\rm Ratio}=\delta \rho^{\rm g}_{yx}/\rho^{\rm T}_{yx}$ with $M=2$meV.
		(b) The magnetization magnitude $M$ dependence of ${\rm Ratio}=\delta \rho^{\rm g}_{yx}/\rho^{\rm T}_{yx}$ with $n=10^{12} {\rm cm}^{-2}$. Both (a) and (b) take the momentum relaxation time is $\tau=1$ ps, effective electron mass $m=0.22m_0$. We have set $l_\phi/l= 5, 10$ in agreement with what is measured in experiment.}
	\end{center}
\end{figure}

Taking dilute magnetic semiconductor as example, a 2D layer containing randomly distributed magnetic skyrmions with sheet density $n_{\rm sk}$ is much lower than the concentration of background nonmagnetic impurities.  In the dilute regime of low skyrmion surface density considered, the total transport scattering time is independent of the carrier spin, being determined by scattering on host nonmagnetic impurities\cite{Denisov-Nonadiabati-PRL-2016}. The substantial advantage of dilute magnetic semiconductors is that both the Fermi energy and the exchange interaction strength can be tuned, allowing us to control the adiabatic parameter in a wide range covering the weak coupling and adiabatic regimes of topological Hall effect. For electrons, the exchange interaction $M=2$ $meV$, the concentration $n=10^{11}$ ${\rm cm}^{-2}$, $m=0.22$ $m_0$, the momentum relaxation time $\tau=1$ ps, we can get the adiabatic parameter $M\tau/\hbar=3$ and the momentum scattering regime $\hbar/\epsilon_{\rm F}\tau=0.6$. The condition of adiabaticity is fulfilled. Here we plot the carrier density and band splitting dependencies of ${\rm Ratio}=\delta \rho^{\rm g}_{yx}/\rho^{\rm T}_{yx}$ in figure~\ref{Fig3}(a) and (b). The ratio between the weak localization correction to the topological Hall resistivity and topological Hall resistivity is around $10^{-3}$, but up to $10^{-2}$ for small $M=1$ $meV$ in diluted magnetic semiconductors. In addition, typical measured value of the giant topological Hall resistivity in FeGe thin films is 80 n$\Omega$ cm to 160 n$\Omega$ cm \cite{THE-exp-PRB-2021,PhysRevLett.108.267201}. Therefore, the weak localization correction to topological Hall resistivity is expected to be observable in these systems.

In addition to the above, a large topological Hall effect occurs in single layer SrRuO$_3$ \cite{THE-SrRuO3} and interface-driven SrRuO$_3$-SrIrO$_3$ superlattice \cite{THE-superlatice}. The maximum of the topological Hall resistivity in these materials can be 0.4 $\mu\Omega{\rm cm}$. Topological insulators interfaced with a magnetic insulators have also reported a topological Hall resistivity around 0.2 $\mu\Omega {\rm cm}$ \cite{THE-TI-interface}, with 1.5 $\mu\Omega {\rm cm}$ in CrTe$_2$/Bi$_2$Te$_3$ \cite{NL-THE-TI}. In addition, in the 2D antiferromagnetic Topological Insulator MnBi$_4$Te$_7$, the topological Hall  resistivity reaches a giant value of 7 $\mu\Omega {\rm cm}$ \cite{MnBi4Te7-Giant-THE}. The Fermi level and band gap can be tuned for MnBi$_4$Te$_7$ by defect engineering and chemical doping in experiment \cite{MnBi4Te7-Tune-EF}. In light of the above we expect an observable value for the weak localization correction to the topological Hall resistivity in experimental measurements of the 2D antiferromagnetic topological insulator MnBi$_4$Te$_7$.

In contrast to the dilute arrays of spin textures mentioned above, the theory is also applicable to skyrmion crystals, which are characterized by a higher density of magnetic vortices $n_{\rm sk} a^2<1$ with $n_{\rm sk}$ the skymions density. This assumption is typical when considering electron transport in real skyrmion crystals. An electron spin state is co-aligned everywhere with a local magnetization and skyrmions do not lead to spin-flip scattering, satisfying the strong exchange interaction we consider in the paper. Nevertheless, further experimental data on these materials is needed before a reliable estimate of weak localization in Hall transport can be produced.

\section{Conclusion}\label{Conc}

We have demonstrated that a quantum correction to the topological Hall effect existst in general and we have provided a comprehensive theoretical blueprint for evaluating it by developing a full theory of the quantum interference-induced corrections to transverse charge transport. Our theory is applicable to a number of systems that host inhomogenous chiral spin textures, and we have estimated the correction for the different materials with large topological Hall resistivity. The theoretical results discussed in this paper will stimulate experiments on topological Hall effect measurement and fill the knowledge gap on the weak localization correction to the transverse transport. Further attention must be awarded to distinguishing the topological Hall effects from other effects that may be present in experiment. From an experimental point of view it is still challenging to differentiate the topological contribution from anomalous Hall effect as along with the topological Hall effect the formation of magnetic skyrmions also reduces an average magnetization and, hence, the anomalous Hall effect \cite{Challenge-THE-review}.

\section*{Acknowledgement}
This work is supported by the Australian Research Council Centre of Excellence in Future Low-Energy Electronics Technologies, project number CE170100039.

\section*{Data availability statement}
All data that support the findings of this study are included within the article (and any supplementary files).
\section*{References}
\bibliographystyle{iopart-num}
%\bibliography{ref-WL-THE}
\providecommand{\newblock}{}

\end{document}